\documentclass[aps,prd,onecolumn,amssymb]{revtex4}
\usepackage{graphicx,bm,color}
\usepackage{amsmath}
\usepackage{amssymb}
\usepackage{amsfonts}
\usepackage{epsfig}
\newcommand{\be}{\begin{equation}}
\newcommand{\ee}{\end{equation}}
\newcommand{\bea}{\begin{eqnarray}}
\newcommand{\eea}{\end{eqnarray}}
\newcommand{\beaa}{\begin{eqnarray*}}
\newcommand{\eeaa}{\end{eqnarray*}}

\allowdisplaybreaks[4]

\begin{document}

\title{Non-locally Reconstructed both Sides of Wormhole and its Non-Traversability }
\author{H. Hadi$^1$}\email{hamedhadi1388@gmail.com}
\affiliation{ $^1$Faculty of Physics, University of Tabriz, Tabriz 51666-16471, Iran}

\begin{abstract}
The theoretical implications of a traversable wormhole between entangled black holes are significant in terms of non-locality and superluminal signaling. By utilizing the entangled states of two maximally entangled black holes, it becomes possible to construct a wormhole. In this context, the traversable wormhole can be achieved by exciting the vacuum state of the near-horizon region of one black hole, allowing for the transmission of information to an observer situated in the near-horizon region of the other black hole through the wormhole. However, the occurrence of this phenomenon is restricted when the near-horizon regions of the black holes undergo a non-local reconstruction, which is referred to as the $ER=EPR$ correspondence. We argue that this correspondence imposes a fundamental limitation on the potential traversability of wormhole and the occurrence of superluminal signaling.
\end{abstract}
\maketitle

\section{Introduction}

The concept of wormholes emerged in the field of modern physics shortly after the discovery of black holes\cite{a1,a2}. It took several decades to fully comprehend the intricate nature of both wormholes and black holes and to recognize their potential significance in the natural world. While there is mounting evidence supporting the existence of astrophysical black holes, the existence of Lorentzian and traversable wormholes remains primarily a fascinating possibility, albeit one with potential observational implications \cite{a3}. One fundamental distinction between these two types of solutions lies in the energy momentum that supports their respective geometries. Black holes exist in a vacuum as the result of gravitational collapse, whereas traversable wormholes necessitate a matter content that violates the null energy condition \cite{a4,a5}. When considering a classical setting and a field theory source, the matter fields (bosonic matter fields) associated with traversable wormholes must possess a nonstandard Lagrangian, such as "phantom" fields \cite{a6,a61,a62,a63,a64}, or one must explore gravity extensions beyond the realm of general relativity \cite{a7,a8}.

To avoid the need for exotic matter in achieving traversable wormholes, alternative approaches involving generalized theories of gravity are explored. Among these, higher-curvature theories of gravity offer promising possibilities for the existence of stable wormholes. Notably, the low-energy heterotic string effective theory \cite{b7,b8} has established a framework for such a generalized gravitational theory in four dimensions. In this theory, the traditional curvature term R from Einstein's theory is augmented by the inclusion of additional fields and higher-curvature gravitational terms. In this particular scenario, additional endeavors have been undertaken, which do not require any type of unusual material. These endeavors involve the exploration of traversable wormholes within the framework of dilatonic Einstein-Gauss-Bonnet theory in four dimensions of spacetime \cite{a7}. Furthermore, recent research has focused on utilizing fermions and linear electromagnetic fields as sources for a wormhole solution, aiming to enable its traversal \cite{prl}.

The aforementioned attempts and solutions primarily focus on situations where wormhole solutions exist independently of any black hole. However, within the framework of general relativity, there are solutions that connect two distant black holes through a wormhole, also known as an Einstein-Rosen bridge. These solutions can be understood as highly entangled states of two black holes, forming a complex $EPR$ pair\cite{cool}. In this study, we specifically examine the traversability of this type of wormhole solution. To do so, it is crucial to understand the behavior of the near horizon region of a black hole in general, or the near horizon region of two entangled black holes in particular. Therefore, it is necessary to review the quantum effects of the near horizon region of black holes and some characteristics of black holes themselves for the purpose of this study. One of the intriguing concepts in black hole physics is the concept of complementarity.

The concept of complementarity  in relation to black holes has faced a challenge from the firewall concept. The $ER = EPR$ conjecture\cite{cool}, which aims to preserve complementarity, does not align with the AMPS proposal of a firewall at the black hole's horizon to resolve the paradox presented by APMS \cite{1}. AMPS argues that accepting all three assumptions simultaneously leads to a contradiction. These assumptions include the preservation of quantum information during the evaporation of a black hole, the non-unusual nature of the event horizon for an observer crossing it, and the compatibility of an observer outside the black hole with relativistic effective quantum field theory. AMPS considers that the late radiation $B$ emitted by an old black hole \footnote{An old black hole is referred to as a black hole that has emitted over fifty percent of its original entropy during the Page time. The concept of an old black hole involves the collapse of a pure state followed by the emission of Hawking radiation, which can be categorized into early and late components denoted as $|\Psi\rangle= \sum_{i}|\psi\rangle_{E}\otimes|i\rangle_{L}$.} (emitted half of its radiation away \cite{2,3}) is maximally entangled with its early radiation $R_B$.The necessity for assumptions 1 and 3 is that the $B$ must be entangled with a subsystem of $R$, while assumption 2 results in entanglement between $B$ and a subsystem within the black hole. This situation contradicts the principle of monogamy in quantum entanglement which states that two maximally entangled systems cannot be entangled with a third system \cite{4,5}. To resolve this contradiction, AMPS proposes that there is only one singularity at the firewall and that the interior of the black hole does not exist \cite{1,6}.

The $ER = EPR$ conjecture is proposed as one of the solutions to overcome AMPS's paradox near the horizon without violating the equivalence principle. The $ER$ bridge and the $EPR$ pair are related through the equation $ER = EPR$ \cite{cool}. This implies that the $ER$ bridge is formed by the $EPR$ correlation in the microstates of two entangled black holes. This finding is supported by the works \cite{24,25}. To elaborate further, the $EPR$ correlated quantum system can be described as a weakly coupled Einstein gravity. In simpler terms, the $ER$ bridge is a highly quantum object. There are speculations suggesting that there is a quantum bridge of this nature for every singlet state. For a more in-depth discussion of AMPS's paradox and an alternative solution, please refer to \cite{26}.

However, the acceptance of a particular and unconventional property within the near horizon region of an older black hole gives rise to a violation of the equivalence principle when considering the $ER = EPR$ conjecture. This violation is referred to as the "zone" \footnote{The concept of the "zone" pertains to a hypothetical boundary situated at a distance roughly equivalent to the Schwarzschild radius from the event horizon of a black hole. Within this paper, we employ the terms "zone" and "near-horizon region" interchangeably.}. Within this zone, the vacuum remains frozen and unaffected by any attempts made by an observer, whether they are in-falling or static, to stimulate it. This distinct behavior of the near-horizon region contradicts the fundamental principles of general relativity, thereby posing a challenge to the equivalence principle.

The current study explores the potential of traversing a wormhole that connects two entangled black holes, with a specific focus on the unique characteristics of the near horizon region of these black holes. As a result, section II presents a comprehensive review and analysis of the distinct features exhibited by the two entangled black holes. Section III is dedicated to investigating the peculiar behavior observed in the near horizon region of the two entangled black holes that are connected by a wormhole, including the examination of the possibility of superluminal signaling through the wormhole. Section IV delves into the exploration of the possibility and impossibility of transmitting information via the wormhole. Finally, a conclusion section summarizes the findings.

\section{Two entangled black holes}
\begin{figure}
    \centering
    \includegraphics[width=0.4\linewidth]{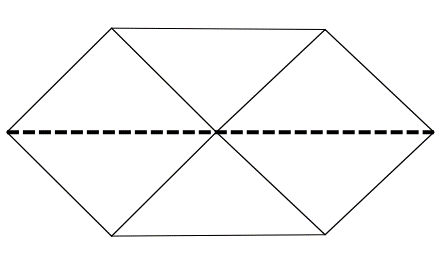}
    \caption{Two-sided Schwarzschild black hole with Penrose diagram constructing by maximally extended Schwarzschild spacetime. The bold dashed spatial slice contains the Einstein-Rosen bridge connecting two asymptotic regions}
    \label{fig:SWCH}
\end{figure}

In this segment, we present several methods for constructing entangled black holes that are connected by a wormhole. The maximally extended Schwarzschild spacetime can be understood as a pair of entangled black holes $(\ref{fig:SWCH})$. These entangled black holes do not possess firewalls, and individuals residing in either the left or right black hole do not experience any peculiar sensations when crossing the event horizon. We can envision two observers inhabiting opposite sides of the Penrose diagram $(\ref{fig:SWCH})$. In our hypothesis, Alice resides on the left sheet while Bob resides on the right sheet. Both spatial sheets are asymptotically flat and possess identical black holes. The bold dashed line represents the initial state, depicting the state as described
\begin{equation}\label{entangled}
    |\Psi\rangle =\sum_{n}e^{-\beta E_{n}/2}|n\rangle |n\rangle.
\end{equation}
Here, $\beta$ is the inverse temperature of the black hole. The tensor product state $|n\rangle_{L} \otimes |m\rangle_{R}$, which represents the entanglement between two systems, can be conveniently denoted as $|n\rangle |n\rangle$. The entanglement is manifested by the fact that the two bifurcate horizons come into contact at the origin. At $t=0$, the spatial geometry resembles a wormhole that connects two separate sheets, each causally disconnected from the other. Despite the geometric connection between these regions, information cannot be transmitted through the bridge due to their causal disconnection. The entanglement of two black holes can be understood in terms of an Einstein-Rosen bridge, which links their geometries across distinct and non-interacting worlds. This entanglement is visualized by matching the bifurcate horizons of the black holes and extending the space-time to include the interior regions behind these horizons. The propagation of non-local signals through the bridge is consistent with the fact that entanglement does not imply non-local signal propagation. The resemblance between the behavior of the Einstein-Rosen bridge and the concept of ${\rm EPR}$ is noteworthy.

\begin{figure}
    \centering
    \includegraphics[width=0.49\linewidth]{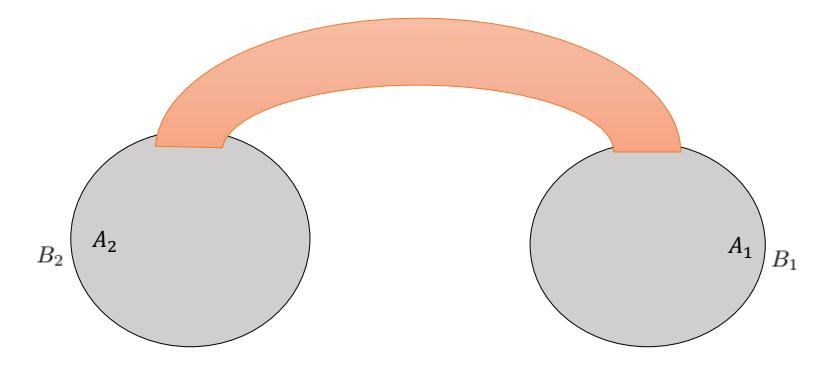}
    \caption{Two-entangled black holes which are connected by a wormhole. Here the two black holes are in identical spacetime and  they are causally connected.}
    \label{fig:2entangled}
\end{figure}

Now we can apply the ${\rm ER=EPR}$ correspondence to a physical model that is not commonly associated with the eternal Schwarzschild black hole. Instead of considering two separate sheets with black holes, we can alternatively examine two extremely distant black holes within a single space. Please refer to figure $(\ref{fig:2entangled})$. The lack of entanglement between these black holes suggests that there is no Einstein-Rosen bridge connecting them. However, if these black holes were initially created in an entangled state $(\ref{entangled})$ at $t=0$, then the bridge would represent their entanglement. Despite the large spatial separation between the two black holes, observers positioned near the horizons of their respective black holes would still be relatively close to each other at least at $t=0$. It is important to note that the natural production of entangled black holes within the same spacetime is theoretically possible, and for further information, one can refer to \cite{cool}.

In this paper, we consider the scenario where an observer named Bob is situated in the vicinity of the first black hole on the right side, while another observer named Alice is positioned just outside the second black hole on the left side. It is important to note that throughout this article, the terms "right/left black hole" and "first/second black hole" are used interchangeably.  

\section{Exciting the vacuum of near horizon region and consequence of it for two connected entangled black holes by an Einstein-Rosen bridge }

In the preceding section, two intertwined concepts of black holes were examined in two distinct models: the maximally extended eternal Schwarzschild black hole and two distant entangled black holes. In the latter scenario, let us consider Bob, who is situated in close proximity to the first black hole, and Alice, who is located near the second black hole. The state of the outer region of Bob's black hole is denoted as $|n\rangle_{B_{1}}$, where $B_{1}$ represents the region adjacent to the horizon of Bob's black hole. The inner region of Bob's black hole is labeled as $A_{1}$ and possesses the state $|n\rangle_{A_{1}}$. The regions $A_{1}$ and $B_{1}$ are entangled, which can be defined as follows
\begin{equation}\label{firstvacuum}
    |\psi \rangle_{A_{1}B_{1}} := \sum_{n=1}^{n}e^{x_{n}}|n\rangle_{A_{1}}|n\rangle_{B_{1}},
\end{equation}  
where $x_{n}=-\frac{n\beta \omega}{2}$. Let us omit the normalization factors ($\sqrt{1-x^{2}}$) for simplicity. Consider the modes that have a Killing frequency comparable to the Hawking temperature. For these modes, the Boltzmann factor $x^{2} = e^{-\beta \omega}$ is of order $1$. Conversely, the adjacent boundary of Alice's black hole is denoted as $B_{2}$ and possesses the state $|n\rangle_{B_{2}}$. In this scenario, the interior region of her black hole $A_{2}$ is entangled with the external region $B_{2}$, and it is characterized by the state $|n\rangle_{A_{2}}$. This entanglement is represented by the equation: 
\begin{equation}\label{secondvacuum}
    |\psi \rangle_{A_{2}B_{2}} := \sum_{n=1}^{n}e^{x_{n}}|n\rangle_{A_{2}}|n\rangle_{B_{2}}
\end{equation}

Under suitable circumstances, Bob and Alice have the ability to enter their respective black holes and utilize the ${\rm ER}$ bridge to encounter one another within. Nevertheless, Bob is unable to meet Alice by jumping into his black hole and accessing the wormhole on the opposite side of her black hole. It is important to note that they can still meet each other outside their black holes by embarking on an extensive journey through spacetime, traversing a considerable distance.

Based on our understanding of entangled black holes, it is evident that the states within the right black hole, denoted as $|n\rangle_{A_{1}}$, are entangled with the internal states of the second black hole, represented as $|n\rangle_{A_{2}}$. This implies that there exists a correlation between $A_{1}$ and $A_{2}$, which can be explained through the ${\rm ER=EPR}$ correspondence $(A_{1}=A_{2})$. This correspondence establishes a mapping between the internal state of the first black hole and the internal state of the second black hole. Furthermore, it is important to acknowledge that Bob possesses access to the near horizon region of his black hole, characterized by the state $(\ref{firstvacuum})$, while Alice has access to the region defined by equation $(\ref{secondvacuum})$.

Now, the impact of Bob or Alice's interaction with the vacuum in the vicinity of the wormhole's horizon can be taken into account when assessing its traversability. The rationale behind considering these interactions stems from the understanding that real physical systems are not isolated, and their interactions with the environment must be considered.The environment could be explicitly incorporated, however, the same outcome can be achieved by regarding the pointer as the environment. By excluding the environment from the depiction, the system will enter a state of mixture, resulting in the collapse of the wave function.

To accomplish this, a pointer $p$ can be introduced, which remains non-interacting with either of the subsystems initially. Consequently, Bob's black hole in the vacuum which is not interacting yet with the pointer $p$ state can be expressed as
\begin{equation}
     |\psi \rangle_{A_{1}B_{1}p}=|0\rangle_{p} \otimes \sum_{n=1}^{n}e^{x_{n}}|n\rangle_{A_{1}}|n\rangle_{B_{1}}
\end{equation}
A possible expression for the vacuum state of Alice's black hole, with a non-interacting pointer, is
\begin{equation}
     |\psi \rangle_{A_{2}B_{2}p}=|0\rangle_{p} \otimes \sum_{n=1}^{n}e^{x_{n}}|n\rangle_{A_{2}}|n\rangle_{B_{2}}
\end{equation}
Suppose Bob possesses knowledge about the state of $B_{1}$ and approaches the vicinity of Alice's black hole, where he encounters Alice. In this encounter, Bob gains an understanding of the state of the nearby region known as $B_{2}$, as he is already acquainted with the internal states of Alice's black hole, denoted as $A_{2}$. This understanding is supported by the concept of $ER=EPR$, which asserts that $A_{2}$ is equivalent to $A_{1}$. In addition, Bob knows that $B_{2}$ is purified by $A_{2}$.

\begin{figure}
    \centering
    \includegraphics[width=0.43\linewidth]{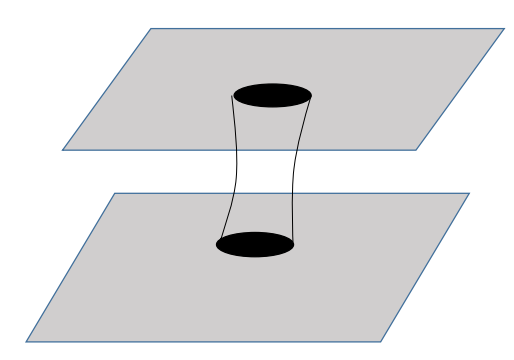}
    \includegraphics[width=0.43\linewidth]{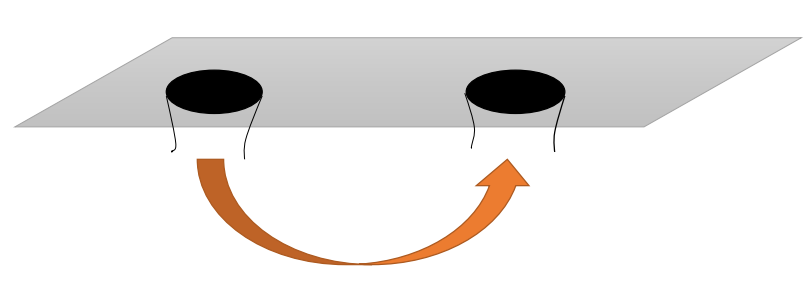}
    \caption{Left:Two-entangled black holes which are connected by a wormhole and they are in two separate spacetime like as figure $(\ref{fig:SWCH})$ Right: Another illustration of figure $(\ref{fig:2entangled})$ where two black holes are in identical spacetime and they are causally connected.}
    \label{fig:3}
\end{figure}

Hence, Bob possesses information regarding the entanglement of two black holes, and he is cognizant of the fact that $B_{2}$ is entangled with $A_{2}$, as he is aware of the entangled component $A_{2}$ of $B_{2}$ derived from his own black hole. However, a question arises as to what will occur if Bob encounters Alice, who has conducted a measurement on $B_{2}$ using a pointer $p$, prior to Bob's arrival in the $B_{2}$ vicinity.
\begin{equation}\label{eq1}
     |\psi \rangle_{A_{2}B_{2}p}=  \sum_{n=1}^{n}e^{x_{n}}|n\rangle_{A_{2}}|n\rangle_{B_{2}}|n\rangle_{p}.
\end{equation}
 Alice can also measures $A_{2}$ instead of $B_{2}$ by falling into her black hole. In this case  $|n\rangle_{p}$ can be written with $|n\rangle_{A_{2}}$ as follows  
\begin{equation}\label{exitedmap}
     |\psi \rangle_{A_{2}B_{2}p}=  \sum_{n=1}^{n}e^{x_{n}}|n\rangle_{A_{2}}|n\rangle_{p}|n\rangle_{B_{2}}.
\end{equation}
This interaction also results in state $(\ref{eq1})$. In the subsequent discussion, we consider the scenario where Alice and her vacuum, after being stimulated, find themselves in the state $(\ref{eq1})$. Under these circumstances, the internal state $A_{2}$ can be understood as the map provided below.
\begin{equation}\label{map}
  |n\rangle_{A_{1}} \longrightarrow   |n\rangle_{A_{2}}  
\end{equation}
However, if he encounters Alice, Bob anticipates being able to identify the state of $B_{2}$ as the state that is purified by $|n\rangle_{A_{2}}$, which he is familiar with from the map $(\ref{map})$, and due to the fact that his vacuum region $B_{1}$ is entangled with $A_{1}$. Conversely, based on Alice's report, he is aware that the state entangled with region $A_{2}$ is $|n\rangle_{B_{2}}|n\rangle_{p}$.

Hence, a crucial query arises: What does Bob consider as the entangled component of the state $|n\rangle_{A_{2}}$ based on his perspective? If Bob does not encounter Alice, he will acknowledge that the entangled part of the state $|n\rangle_{A_{2}}$ is $|n\rangle_{B_{2}}$, and he can determine this state through the purification of $|n\rangle_{A_{2}}$. However, what will occur if Bob meets Alice after she has performed a measurement using a pointer $p$ on $B_{2}$ as shown in equation $(\ref{eq1})$? In this scenario, Bob will face a contradiction. On one hand, he knows that the entangled part of the state $|n\rangle_{A_{2}}$ is $|n\rangle_{B_{2}}|n\rangle_{p}$, but on the other hand, without encountering Alice, he knows from his understanding of entangled black holes that the entangled part of $|n\rangle_{A_{2}}$ is $|n\rangle_{B_{2}}$. It should be noted that Bob always knows the state $|n\rangle_{A_{2}}$ based on his knowledge of black holes and the map $(\ref{map})$.

It should be noted that the state $|n\rangle_{A_{2}}$ cannot be purified simultaneously with both $|n\rangle_{B_{2}}$ and $|n\rangle_{B_{2}}|n\rangle_{p}$, unless Bob becomes aware of Alice's measurement from her vacuum of the near horizon region $B_{2}$, as defined by equation $(\ref{eq1})$, in his own vacuum prior to commencing his journey. In simpler terms, when Alice stimulates her vacuum, Bob must be able to detect this stimulation in his own vacuum. Consequently, Bob will observe his vacuum state as $|n\rangle_{B_{1}}|n\rangle_{p}$ instead of just $|n\rangle_{B_{1}}$ following Alice's measurement $(\ref{eq1})$. This leads to a superluminal signaling through the wormhole or traversability of wormhole. In the subsequent section, this phenomenon will be critically examined and explored in greater detail.

\section{Transforming information through the wormhole}
This section examines the analysis of the two interlinked black holes and the possibility of traversing the Einstein-Rosen wormhole in two scenarios. Initially, the investigation focuses on the situation where the two black holes are located in causally connected sheets, resembling the right shape depicted in figure $(\ref{fig:3})$. Subsequently, the scrutiny extends to the case where the two entangled black holes exist in separate sheets, resembling either the left shape shown in figure $(\ref{fig:3})$ or the figure $(\ref{fig:SWCH})$.

\subsection{Two entangled black holes in the casually connected spacetime}
Assume that Alice and Bob find themselves in the vicinity of their respective black holes, as depicted in figures $(\ref{fig:3})$ (left) or $(\ref{fig:2entangled})$. Although these two black holes are causally linked, they are situated at a considerable distance from each other, thereby remaining unaffected by each other's gravitational attraction. Consequently, both Alice and Bob experience their own gravitational waves and near horizon vacuum. 

In the preceding section, it was mentioned that when Bob encounters Alice after her measurement on the near horizon vacuum, he is faced with a contradiction. Let us now consider a scenario where Alice and Bob, prior to positioning themselves in their respective black hole vacuums, have decided to exchange information with each other after positioning. In order to do so, Alice performs a measurement resulting in the occupation number eigenstate of $B_{2}$ according to equation $(\ref{eq1})$. On the other hand, Bob finds himself either in the vacuum of the right side black hole or on the opposite side of the wormhole that connects the two entangled black holes. He has the option to fall into the black hole and experience the vacuum. However, what would happen if Bob, without conducting any specific measurement, carries a sufficiently sensitive detector to detect any deviations from empty space? Considering that Alice has performed a measurement leading to $B_{2}$ being in an occupation number eigenstate, Bob should observe a deviation from empty space. Prior to Alice's measurement, based on his knowledge of black holes and the $ER=EPR$ correspondence, Bob knows that his vacuum is defined by equation $(\ref{firstvacuum})$ and he expects it to have no deviations from empty space when using a sensitive enough detector. However, after Alice's measurement, the $ER=EPR$ correspondence implies a different mapping as follows
\begin{equation}\label{map2}
  |PSB\rangle \longrightarrow |n\rangle_{A_{1}}, 
\end{equation}
where $|PSB\rangle$ is the purified state of $|n\rangle_{B_{2}}|n\rangle_{p}$. It is important to bear in mind that the purified state of $|n\rangle_{A_{2}}$ is equivalent to $|n\rangle_{B_{2}}$, but this only occurs when Alice avoids from exciting her black hole's vacuum. On the other hand, Bob, who is aware that the outside state $B_{1}$ of his black hole is purified by the inside state $A_{1}$, possesses complete knowledge of his vacuum $(\ref{firstvacuum})$. However, when Alice performs a measurement on her vacuum $(\ref{eq1})$, the map $(\ref{map2})$ is applied, resulting in Bob discovering that the purified state of $|n\rangle_{B_{1}}$ is not $|n\rangle_{A_{1}}$ or even $|n\rangle_{A_{2}}$, but rather $|PSB\rangle$ in accordance with the $ER=EPR$ correspondence. This discrepancy implies that Bob must recognize a different state for the outside region $B_{1}$ that is distinct from $|n\rangle_{B_{1}}$. Consequently, Alice's act of exciting her vacuum exerts an influence on Bob's vacuum in the outside region $A_{1}$, ultimately leading to a superluminal signaling through the Einstein-Rosen bridge. This intriguing outcome warrants further examination, which I will delve into with greater scrutiny in the subsequent analysis.

In the scenario where two entangled black holes exist within a causally connected spacetime, an intriguing situation arises. Let us consider Alice and Bob, who find themselves in the near horizon region of their respective black holes, as depicted in figures $(\ref{fig:3})$ (right) or $(\ref{fig:SWCH})$. It is important to note that these two black holes are causally disconnected, resulting in each of them possessing their own gravitational wave and near horizon vacuum.

In the event that an individual replicates the identical thought experiment as the preceding subsection in the asymptotic area of a wormhole, it becomes evident that transmitting information from one side of the wormhole to the other is unattainable. This is due to the fact that the near horizon region of the wormhole is causally disconnected, as depicted by the dashed line in figure $(\ref{fig:SWCH})$. The crucial inquiry that arises is: what specific phenomenon hinders the non-transformation of information through the wormhole or the occurrence of superluminal signaling? The subsequent subsection is dedicated to exploring this very question.

\subsection{Superluminal signaling and violation of equivalence principle of general relativity}
The concept of the Equivalence principle holds significant importance within the framework of general relativity. It states that an observer falling into a black hole would not perceive any distinct changes while crossing the event horizon. Consequently, the stimulation of the near horizon vacuum would not appear unusual to the observer in this particular context. However, as discussed in the preceding subsection, this phenomenon can lead to either a superluminal signaling through the wormhole or the leakage of information between two regions that are causally disconnected, with the wormhole connecting these regions.

To resolve this inconsistency, we may postulate that the near horizon region has a distinctive vacuum state, as suggested by \cite{bosso}.
\begin{equation}\label{frozenvacuum}
    |0 \rangle_{AB} := \sum_{n=1}^{n}e^{x_{n}}|n\rangle_{A}|n\rangle_{B}.
\end{equation}
 In this context, the stationary observer positioned close to the horizon area or an observer falling into the horizon will identify their vacuum state as peculiar and perpetually perceive it as equation $(\ref{frozenvacuum})$. Within this vicinity, there exists no potential for particle creation or stimulation of the vacuum. The frozen vacuum emerges as a consequence of the non-local reconstruction of the vacuum near the horizon region. This non-local reconstruction occurring in the near horizon region ultimately results in the violation of the equivalence principle of general relativity.

The frozen vacuum phenomenon prevents Alice and Bob, who are separate observers on either side of the wormhole, from communicating with each other without falling into their respective black holes. Moreover, the frozen vacuum also hinders communication across the two causally disconnected regions or through a wormhole that links two entangled black holes. Thus, in order to preserve causality and non-local reconstruction of the horizon region, we have to accept the breakdown of the equivalence principle of general relativity.

It should be emphasized that the aforementioned description does not serve as an argument to establish the existence of a frozen vacuum; rather, it confirms its presence. The primary evidence supporting this notion can be found in \cite{bosso}, where an attempt is made to resolve a contradiction in the evaporation process of an old black hole by introducing the concept of a frozen vacuum within the vicinity of its event horizon. However, when considering the model of two entangled black holes, the assumption of this peculiar behavior for the vacuum near the event horizon proves beneficial in avoiding contradictions and preventing the occurrence of superluminal signaling through the wormhole, thereby ensuring the non-traversability of the wormhole solution in the context of the two entangled black holes model.

\section{Conclusion}
In this paper, we analyzed the essential properties of a wormhole that connects two entangled black holes and explored the conditions for its traversability. We indicated that the possibility of superluminal signaling through the wormhole is determined by the asymptotic behaviour of the wormhole near the horizon region of the two entangled black holes. The non-local reconstruction of the near horizon vacuum results in a frozen vacuum that forbids any excitation or particle creation in this region, violating the equivalence principle of general relativity. This phenomenon also prevents communication and superluminal signaling between two separate observers on either side of the wormhole. Furthermore, the frozen vacuum impedes communication across the two causally disconnected regions or through a wormhole that connects two entangled black holes. Therefore, we conclude that if we adhere to the $ER=EPR$ correspondence, we have to sacrifice the equivalence principle of general relativity in order to maintain causality, non-superluminal signaling and non-local reconstruction of the horizon region.


\end{document}